\documentclass[onecolumn,aps,pr,superscriptaddress,preprintnumbers,nofootinbib,10pt]{revtex4-2}
\usepackage{amsmath,amssymb}
\usepackage[dvipdf,dvips]{graphicx}
\usepackage{color}
\usepackage{hyperref}
\usepackage{url}
\usepackage{slashed}
\usepackage{subfigure}
\usepackage{amsmath}
\usepackage{amsfonts}
\usepackage{float} 
\usepackage{amssymb}
\usepackage{epsfig}
\usepackage{graphics}
\usepackage{euscript}
\usepackage{slashed}
\usepackage{epstopdf}
\usepackage[utf8]{inputenc}
\allowdisplaybreaks
\usepackage{pifont}
\usepackage{dsfont}
\usepackage{MnSymbol}
\usepackage{verbatim}
\usepackage{graphicx}
\usepackage{latexsym}
\usepackage{courier}
\usepackage{multirow}
\usepackage{physics}
\usepackage{tikz-feynman}
\usepackage{setspace}

\def \and{\textmd{and}}

\def \x{\vb*{x}}
\def \y{\vb*{y}}

\graphicspath{{./}}
\newbox{\ORCIDicon}
\sbox{\ORCIDicon}{\large\includegraphics[width=0.8em]{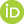}}

\begin{document}

	\title{Bell-CHSH inequality and unitary transformations\\in Quantum Field Theory}
	
	\author{D. O. R. Azevedo} \email{daniel.azevedo@ufv.br} \affiliation{Departamento de Física, Universidade Federal de Vi{\c c}osa (UFV), Avenida Peter Henry Rolfs s/n - 36570-900, Vi{\c c}osa, MG, Brazil} \affiliation{Instituto de Física, Facultad de Ingeniería, Universidad de la República, J. H. y Reissig 565, 11000 Montevideo, Uruguay}
	
	\author{F. M. Guedes} \email{guedes.fillipe@posgraduacao.uerj.br} \affiliation{UERJ $–$ Universidade do Estado do Rio de Janeiro,	Instituto de Física $–$ Departamento de Física Teórica $–$ Rua São Francisco Xavier 524, 20550-013, Maracanã, Rio de Janeiro, Brazil}

	\author{M. S.  Guimaraes}\email{msguimaraes@uerj.br} \affiliation{UERJ $–$ Universidade do Estado do Rio de Janeiro,	Instituto de Física $–$ Departamento de Física Teórica $–$ Rua São Francisco Xavier 524, 20550-013, Maracanã, Rio de Janeiro, Brazil}
	
	\author{I. Roditi} \email{roditi@cbpf.br} \affiliation{CBPF $-$ Centro Brasileiro de Pesquisas Físicas, Rua Dr. Xavier Sigaud 150, 22290-180, Rio de Janeiro, Brazil } 
	
	\author{S. P. Sorella} \email{silvio.sorella@fis.uerj.br} \affiliation{UERJ $–$ Universidade do Estado do Rio de Janeiro,	Instituto de Física $–$ Departamento de Física Teórica $–$ Rua São Francisco Xavier 524, 20550-013, Maracanã, Rio de Janeiro, Brazil}

	\author{A. F. Vieira\,\href{https://orcid.org/0000-0003-2897-2437}{\usebox{\ORCIDicon}}} \email{arthurfvieira@if.ufrj.br} \affiliation{UFRJ $-$ Universidade Federal do Rio de Janeiro, Instituto de Física, RJ 21.941-972, Brazil}

	\begin{abstract}
	Unitary transformations are employed to enhance the violations of the Bell-CHSH inequality in relativistic Quantum Field Theory. The case of the scalar field in $1+1$ Minkowski space-time is scrutinized by relying on the Tomita-Takesaki modular theory. The example of the bounded Hermitian operator $sign(\varphi(f))$, where $\varphi(f)$ stands for the smeared scalar field, is worked out. It is shown  that unitary deformations enable for violations  of  the Bell-CHSH inequality. The setup is generalized to the Proca vector field by means of its equivalence with the scalar theory.
	\end{abstract}

	\maketitle
	
	%%%%%%%%%%%%%%%%%%%%%%%%%%%%%%%%%%%%%%%%%%%%%%%%%%%%%%%%%%%%%%%%%%%%%%%%%%%%%%%%%%%%%%%%%%%	
	\section{Introduction}\label{s1}

The study of Bell inequalities within the framework of quantum theory has been central to elucidating the profound distinction between quantum correlations and those permitted by local hidden variable theories. Originally formulated by Bell in 1964 \cite{Bell:1964kc}, these inequalities place stringent constraints on statistical correlations predicted by any theory respecting locality and realism. Their violation, as predicted by quantum mechanics and experimentally confirmed with increasing sophistication \cite{Clauser:1969ny,Freedman:1972zza,Clauser:1974tg,Clauser:1978ng,Aspect:1976fm,Aspect:1981zz,Aspect:1981nv,Aspect:1982fx,Greenberger:1989tfe,Giustina:2015yza}, provides compelling evidence for the intrinsically nonlocal character of entanglement.

While the formulation and experimental verification of Bell inequalities have traditionally been confined to non-relativistic quantum systems with finite degrees of freedom, their extension to the domain of relativistic Quantum Field Theory (QFT) presents both conceptual and technical challenges. Unlike finite-dimensional Hilbert space settings, QFT inherently accommodates an infinite number of degrees of freedom and is governed by  locality and causality. Fields are operator-valued distributions, and their nontrivial vacuum structure (epitomized by the Reeh-Schlieder theorem \cite{Haag:1992hx}) precludes a naive transposition of quantum mechanical results into field-theoretic contexts.

To rigorously investigate entanglement and nonlocality in relativistic systems, it becomes necessary to employ the mathematical tools of Algebraic Quantum Field Theory (AQFT), wherein local algebras of observables are associated with spacetime regions. Within this formalism, the Tomita-Takesaki modular theory plays a pivotal role, especially in light of Haag duality and the structure of von Neumann algebras generated by local field operators. These tools enable a principled construction of observables localized in spacelike separated regions, allowing for the definition of Bell-type inequalities directly in QFT. In a series of theorems, Summers and Werner \cite{Summers:1987fn,Summers:1987squ,Summers:1987ze} have shown that violations of the Bell-Clauser-Horne-Shimony-Holt (Bell-CHSH) inequality are not only possible in QFT but, remarkably, are generic even in the vacuum state, achieving maximal violations already for free fields.\\\\Nevertheless, it is safe to state that a concrete example of a field theory model implementing in a computational setup the above mentioned theorems is still lacking. In recent years, we have devoted much efforts to build up such a concrete model. By combining the Tomita-Takesaki modular theory with a numerical approach, we have been able to report  examples of violations of the Bell-CHSH inequality, see \cite{DeFabritiis:2023tkh,DeFabritiis:2024jfy,Guimaraes:2024mmp}. Though, the explicit construction of both field operators and relative test functions leading to the saturation of the Tsirelson bound looks still very challenging. \\\\Let us give a short overview of what we have been able to work out so far. Formulating the Bell-CHSH inequality in QFT requires three key ingredients, namely: the choice of the state, the specification of the geometry of the spacetime regions under investigation and the construction of the Hermitian bounded field operators entering the Bell-CHSH correlator. Our contribution to these issues can be found in \cite{DeFabritiis:2023tkh, DeFabritiis:2024jfy,Guimaraes:2024lqf,Guimaraes:2024xtj}, being reviewed in \cite{Guimaraes:2024mmp}. With respect to the choice of the state, we have focused on the vacuum state. In \cite{DeFabritiis:2024jfy} a numerical study of the Bell-CHSH inequality in the left and right Rindler wedges can be found, while in \cite{Guimaraes:2024xtj} diamond shaped regions are investigated. Parallel to the numerical approaches, the Tomita-Takesaki modular theory has been employed to study the properties of the correlation functions of Weyl operators \cite{DeFabritiis:2023tkh}. The relevance of the modular theory for the investigation of aspects of entanglement in QFT has been emphasized by \cite{Chatterjee:2024cpd,Kasprzak:2024rzj,Chatterjee:2025pky,Tjoa:2021roz}. \\\\More recently, in \cite{Guimaraes:2025vfu} we have addressed the very difficult issue of the construction of a set of bounded Hermitian operators whose correlation functions can be effectively evaluated, through analytic and numerical methods. It is worth to point out that this issue is far from receiving a satisfactory answer. To our knowledge, the explicit construction of the field operators which saturate the Tsirelson bound is far from being achieved. More precisely, as shown in \cite{Guimaraes:2025vfu}, a class of bounded Hermitian localized operators $\{ {\cal F}_f \}$ has been introduced through continuous superpositions of the Weyl operators (see section \ref{QFT}), namely 
\begin{equation} 
	{\cal F}_f = \int_{-\infty}^{\infty} dk \;\rho(k)\; e^{i k \varphi(f)} \;, \label{Fop}
\end{equation}
where $\varphi(f)$ is the smeared field and $\rho(k)$ a suitable normalized distribution. As discussed in \cite{Guimaraes:2025vfu}, this class of operators has enabled a better understanding of several aspects related to the QFT formulation of the Bell-CHSH inequality. \\\\In the present work, we combine the setup of \cite{Guimaraes:2025vfu} with the use of unitary transformations to  improve the size of the violations. It is worth mentioning that the use of unitary transformations  is largely employed in Quantum Mechanics in order to optimize the results of the study of the Bell-CHSH inequality. \\\\As a concrete example, we shall perform a detailed study of the operator ${\it sign}(\varphi(f))$, defined through the Dirichelet representation:
\begin{equation} 
	{\it sign}(\varphi(f)) = \frac{2}{\pi} \int_{0}^{\infty} \frac{dk}{k}\; \sin(k \varphi(f) ) \;. \label{sign}
\end{equation}
As will be demonstrated, this operator is instrumental in clarifying the manner in which unitary transformations influence the Bell-CHSH inequality in QFT. In particular:
\begin{itemize} 
	\item The correlation functions associated with the operator \eqref{sign} admit a closed-form evaluation, which can be expressed in terms of the imaginary error function.
	
	\item In the absence of unitary transformations, the correlation functions derived from \eqref{sign} do not yield violations of the Bell-CHSH bound.
	
	\item The introduction of unitary transformations modifies the structure of the operator \eqref{sign}, generating additional parameters that render violations of the Bell-CHSH inequality possible.
	
\end{itemize}

The paper is organized as follows. In Section \ref{QM}, we review the role of unitary transformations in the Bell-CHSH inequality in Quantum Mechanics. Section \ref{QFT} introduces the modular-theoretic construction of localized operators in QFT and applies it to the real scalar field. Section \ref{ProcaField} extends the analysis to the Proca field in $(1+1)$ dimensions. In higher dimensions, a Proca field propagates massive spin-1 excitations with multiple polarization states. However, in two dimensions the transversality condition eliminates any independent polarization, leaving only a single propagating degree of freedom. More precisely, the Proca field is \emph{dually equivalent} to a massive scalar field: one can represent the field strength as $F_{\mu\nu} = m^2 \,\varepsilon_{\mu\nu}\,\varphi$, or, equivalently, the vector potential as $A^\mu = \varepsilon^{\mu\nu}\partial_\nu \varphi$, where $\varphi$ is a scalar field and $\varepsilon_{\mu\nu}$ is the two-dimensional Levi-Civita symbol.\footnote{This duality is exact at the level of free field theories and ensures that the physical Hilbert spaces of the Proca and scalar models are isomorphic. The massive scalar field may thus be regarded as the unique dynamical degree of freedom encoded in the Proca field in $1+1$.} By carefully implementing the smearing procedure with transverse test functions and exploiting this equivalence, we show that the Bell-CHSH violation in the Proca vacuum reproduces exactly the scalar case,  highlighting how dualities between field theories can directly shape the manifestation of quantum nonlocality. Our conclusions and perspectives for future developments are presented in Section \ref{Cc}.

\section{Unitary transformations and the Bell-CHSH inequality in Quantum Mechanics}\label{QM}

Let us begin by providing a short account of the Bell-CHSH inequality in Quantum Mechanics. For a bipartite system, the Bell-CHSH inequality reads:
\begin{equation} 
	\langle \psi |\; {\cal C}\;| \psi \rangle = \langle \psi |\; (A+A') \otimes B + (A-A')\otimes B'\; |\psi\rangle \;, \label{bcsh}
\end{equation}
where $|\psi\rangle$ stands for the state of the system whose Hilbert space is ${\cal H}= {\cal H}_A \otimes {\cal H}_B$. As it is customary, the letters $(A,B)$ refer to Alice and Bob. A violation of the Bell-CHSH inequality occurs whenever 
\begin{equation} 
	2  <  \Big|\langle \psi |\; {\cal C}\;| \psi \rangle \Big| \le 2 \sqrt{2} \;, \label{vb}
\end{equation}
where the bound $2\sqrt{2} \approx 2.83$ is the so-called Tsirelson’s bound \cite{Cirelson:1980ry}. The Alice's and Bob's operators, $(A,A')$ and  $(B,B')$, are called Bell's observables and act on ${\cal H}_A$ and ${\cal H}_B$, respectively. These operators are Hermitian and dichotomic, fulfilling the conditions:
\begin{eqnarray}
	A & = & A^\dagger \;, \qquad A'= {A'}^\dagger \;, \qquad A^2={\mathbf{1}} \;, \qquad {A'}^2={\mathbf{1}} \;, \nonumber \\
	B & = & B^\dagger \;, \qquad B'= {B'}^\dagger \;, \qquad B^2={\mathbf{1}} \;, \qquad {B'}^2={\mathbf{1}} \;, \label{c1} 
\end{eqnarray} 
and 
\begin{equation}
	[A,B]=[A',B] = [A,B']=[A',B']=0 \;, \qquad [A,A']\neq 0 \;, \qquad [B,B']\neq 0 \;. \label{c2}
\end{equation} 
From eqs.\eqref{c1},\eqref{c2} it is apparent that the Bell operators are defined up to unitary transformations. It is straightforward to check that 
\begin{eqnarray} 
	{\hat A} & = &  {\cal U}^\dagger_A \; A \; {\cal U}_A  \;, \qquad {\hat A'}  =   {\cal U}^\dagger_A \; A' \; {\cal U}_A  \nonumber \\
	{\hat B} & = &  {\cal U}^\dagger_B \; B\; {\cal U}_B  \;, \qquad {\hat B'}  =   {\cal U}^\dagger_B \; B' \; {\cal U}_B \;,\label{uuu}
\end{eqnarray} 
where ${\cal U}_A $ and ${\cal U}_B$ are unitary transformations operators acting on ${\cal H}_A$ and ${\cal H}_B$, fulfill the same relations obeyed by $(A,A')$ and $(B,B')$. This freedom is largely employed to enhance the size of the violations of the Bell-CHSH correlator. \\\\Consider in fact  the entangled spin-$1/2$ singlet state
\begin{equation} 
	| \psi_{AB}\rangle = \frac{1}{\sqrt{2}}\left( |+\rangle_A \otimes |-\rangle_B - |-\rangle_A \otimes |+\rangle_B\right)  \;, \label{sing}
\end{equation}
which gives maximal violation of the Bell-CHSH inequality:
\begin{equation} 
	\Big|\langle \psi_{AB} |\; {\cal C}\;| \psi_{AB} \rangle \Big| = 2 \sqrt{2} \;. \label{ts}
\end{equation}
In this case, the Bell operators may be written as  
\begin{eqnarray} 
	A & = & e^{i\alpha} |-\rangle_A {}_A \langle +| + e^{-i\alpha} |+\rangle_A {}_A \langle -| \;, \qquad A'  =  e^{i\alpha'} |-\rangle_A {}_A \langle +| + e^{-i\alpha'} |+\rangle_A {}_A \langle -| \;, \nonumber \\
	B & = & e^{i\beta} |-\rangle_B {}_B \langle +| + e^{-i\beta} |+\rangle_B {}_B \langle -| \;, \qquad B'  =  e^{i\beta'} |-\rangle_B {}_B \langle +| + e^{-i\beta'} |+\rangle_B {}_B \langle -| \;, \label{op12}
\end{eqnarray}
where the parameters $(\alpha, \alpha', \beta, \beta')$ denote the so-called Bell angles. The operators in Eq.~\eqref{op12} fulfill the conditions \eqref{c1} and \eqref{c2}. A simple   calculation gives 
\begin{equation} 
	\langle \psi_{AB} |\;A\otimes B \;| \psi_{AB} \rangle = - \cos(\alpha-\beta)  \;. \label{ss12}
\end{equation} 
The usual choice for $(\alpha, \alpha', \beta,\beta')$, corresponding to Eq.~\eqref{ts}, is 
\begin{equation}
	\alpha = 0 \;, \qquad \beta = \frac{\pi}{4} \;, \qquad \alpha'= \frac{\pi}{2} \;, \qquad \beta'= - \frac{\pi}{4} \;. \label{angles}
\end{equation}
Looking now at the Bell operators $(A,A')$, $(B,B')$, one recognizes that they can be obtained from a pair of reference Bell operators, $(A_0,B_0)$, upon acting with unitaries. For example, it turns out that 
\begin{equation} 
	A =  {\cal U}^\dagger_A \; A_0 \; {\cal U}_A  \;, \label{u12}
\end{equation} 
where ${\cal U}_A $ stands for the unitary operator 
\begin{equation} 
	{\cal U}_A = e^{-i \alpha |-\rangle_A {}_A\langle -| }  \;, \label{uf12}
\end{equation}
while $A_0$ is the Hermitian dichotomic reference operator 
\begin{equation} 
	A_0 =   |-\rangle_A {}_A \langle +| \;+\;  |+\rangle_A {}_A \langle -| \;, \qquad A_0 = A^\dagger_0 \;, \qquad A_0^2 = {\mathbf{1}}  \;. \label{a0}
\end{equation}
In the same vein, 
\begin{eqnarray} 
	A' & = & e^{i \alpha' |-\rangle_A {}_A\langle -| } \;  A_0  \; e^{-i \alpha' |-\rangle_A {}_A\langle -| } \;, \qquad
	B =  e^{i \beta |-\rangle_B {}_B\langle -| } \;  B_0  \; e^{-i \beta |-\rangle_B {}_B\langle -| } \;, \qquad 
	B'  =  e^{i \beta' |-\rangle_B {}_B\langle -| } \;  B_0  \; e^{-i \beta' |-\rangle_B {}_B\langle -| } \;,
\end{eqnarray}
with
\begin{eqnarray}
	B_0  &=&   |-\rangle_B {}_B \langle +| \;+\;  |+\rangle_B {}_B \langle -| \;, \qquad B_0 = B^\dagger_0 \;, \qquad B_0^2 = {\mathbf{1}} \;. \label{b0}
\end{eqnarray}
This example shows that all parameters $(\alpha, \alpha', \beta, \beta')$, from which the violation of the Bell-CHSH inequality originates, can be encoded into unitaries. \\\\Other examples may be provided, including entangled coherent states, Greenberger-Horne-Zeilinger (GHZ) states, the $N00N$ states and the squeezed state \cite{Guimaraes:2024byw,DeFabritiis:2023bbs}. The above considerations show the relevant role that unitaries play in the violation of the Bell-CHSH inequality in Quantum Mechanics.

\section{Facing the Bell-CHSH in Quantum Field Theory}\label{QFT}	

We are now ready to face the more complex case of the relativistic QFT. We shall rely on the original works by S. J. Summers and R. Werner \cite{Summers:1987fn,Summers:1987squ,Summers:1987ze}, who have established a set of remarkable results concerning the violation of the Bell-CHSH inequality in the vacuum state, see also the recent review \cite{Guimaraes:2024mmp}.

\subsection{The real massive scalar field in $(1+1)$ Minkowski spacetime}

We consider a free real scalar field of mass $m$ in $(1+1)$-dimensional Minkowski spacetime. Its quantized form, expressed via plane-wave decomposition, reads
\begin{equation}\label{qf}
	\varphi(t,x) = \int \! \frac{d k}{2 \pi} \frac{1}{2 \omega_k} \left( e^{-ik_\mu x^\mu} a_k + e^{ik_\mu x^\mu} a^{\dagger}_k \right), 
\end{equation}
where $\omega_k = \sqrt{k^2 + m^2}$ and the annihilation and creation operators satisfy the canonical commutation relations
\begin{align}\label{eq:CCR}
	[a_k, a^{\dagger}_q] &= 2\pi \, 2\omega_k \, \delta(k - q), \\ \nonumber 
	[a_k, a_q] &= [a^{\dagger}_k, a^{\dagger}_q] = 0. 
\end{align}

In a rigorous QFT framework, the scalar field is treated as an operator-valued distribution \cite{Haag:1992hx}, and thus must be smeared out with smooth, compactly supported test functions to yield well-defined operators on the Hilbert space. Given a test function $h(\vb*{x}) \in C_0^\infty(\mathbb{R}^{1,1})$, with $\vb*{x}=(t,x)$, the corresponding smeared field operator is defined as
\begin{align} 
	\varphi(h) = \int \! d^2\vb*{x} \; \varphi(\vb*{x}) h(\vb*{x}).
\end{align}
The Lorentz-invariant vacuum expectation value of two smeared field operators is defined through the two-point Wightman function:
\begin{align} \label{InnerProduct}
	\langle f \vert g \rangle &= \langle 0 \vert \varphi(f) \varphi(g) \vert 0 \rangle =  \frac{i}{2} \Delta_{\rm PJ}(f,g) +  H(f,g),
\end{align}
where $f, g \in C_0^\infty$, and $\Delta_{\text{PJ}}(f,g)$ and $H(f,g)$ denote the smeared Pauli–Jordan and Hadamard distributions, respectively. These are defined by
\begin{align}
	\Delta_{\rm PJ}(f,g) &=  \int \! d^2\vb*{x} d^2\vb*{y} f(\vb*{x}) \Delta_{\rm PJ}(\vb*{x}-\vb*{y}) g(\vb*{y}) \;,  \nonumber \\
	H(f,g) &=  \int \! d^2\vb*{x} d^2\vb*{y} f(\vb*{x}) H(\vb*{x}-\vb*{y}) g(\vb*{y})\;. \label{mint}
\end{align}
with distributional kernels given explicitly by
\begin{eqnarray} 
	\Delta_{\rm PJ}(\vb*{x}) & =&  -\frac{1}{2}\;{\rm sign}(t) \; \theta \left( \lambda(\vb*{x}) \right) \;J_0 \left(m\sqrt{\lambda(\vb*{x})}\right) \;, \nonumber \\
	H(\vb*{x}) & = & -\frac{1}{2}\; \theta \left(\lambda(\vb*{x}) \right )\; Y_0 \left(m\sqrt{\lambda(\vb*{x})}\right)+ \frac{1}{\pi}\;  \theta \left(-\lambda(\vb*{x}) \right)\; K_0\left(m\sqrt{-\lambda(\vb*{x})}\right) \;, \label{PJH}
\end{eqnarray}
where the Lorentz-invariant quantity
\begin{equation}
	\lambda(\vb*{x}) = t^2 - x^2
\end{equation}
determines the causal structure. Here, ($J_0$, $Y_0$, $K_0$) are Bessel functions of the first kind, second kind, and modified second kind, respectively.\\\\
The Pauli-Jordan distribution $\Delta_{\rm PJ}(\vb*{x})$ is Lorentz-invariant and encodes causality, vanishing outside the light cone. Moreover, it is odd under $\vb*{x} \rightarrow -\vb*{x}$, whereas the Hadamard function $H(\vb*{x})$ is even. The commutator of the field operators is thus $\left[\varphi(f), \varphi(g)\right] = i \Delta_{\rm PJ}(f,g)$, ensuring that  $\left[\phi(f), \phi(g)\right] = 0,$ whenever the supports of $f$ and $g$ are spacelike separated. This compactly encodes the principle of microcausality in relativistic field theory.

\subsection{Basics of the Tomita-Takesaki modular theory} 

The Tomita-Takesaki modular theory serves as a powerful framework for analyzing the Bell-CHSH inequality in QFT \cite{Summers:1987fn,Summers:1987squ,Summers:1987ze}. To set the stage, it is worth briefly reviewing some of the basic features of this elegant  theoretical structure. \\\\Let ${\cal O}$ stand for an open region of the Minkowski spacetime and let ${\cal M}({\cal O})$ be the space of test functions with support contained in $\cal O$: 
\begin{equation} 
	{\cal M}({\cal O}) = \{ f; \; {\rm such \; that} \; supp(f) \subseteq {\cal O} \}. \label{MO}
\end{equation}
One  introduces the symplectic complement of ${\cal M}({\cal O})$ as 
\begin{equation} 
	{\cal M'}({\cal O}) = \{ g; \; \Delta_{\rm PJ}(g,f) =0 \;\; \forall f \in {\cal M}({\cal O})\}. \label{MpO}
\end{equation}
In other words, ${\cal M'}({\cal O})$ is given by the set of all test functions for which the smeared Pauli-Jordan expression defined by Eq.~\eqref{mint} vanishes. The usefulness of the symplectic complement ${\cal M'}({\cal O})$ relies on the fact that it allows us to rephrase causality as 
\begin{equation} 
	\left[ \varphi(g) , \varphi(f) \right] = 0 \;, \;\; \forall \;\;g\in {\cal M'}({\cal O}) \;\;{\rm and} \;\;f \in {\cal M}({\cal O}). \label{MMM}
\end{equation}
The next step is that of introducing a von Neumann algebra  ${\cal W}({\cal M})$ of bounded operators supported in ${\cal O}$, equipped with a cyclic and separating vector, which will be addressed in the discussion of von Neumann algebras later. As a concrete example of such an algebra, one may consider the von Neumann algebra generated by the Weyl operators, namely 
\begin{equation} 
	{\cal W}({\cal M}) = {\rm span}\;\left\{  {\cal A}_h \;, supp(h) \in {\cal M} \right\} \;, \label{vn}
\end{equation}  
where ${\cal A}_h$ stands for the unitary Weyl operator ${\cal A}_h = e^{i {\varphi}(h) }.$
%\begin{equation}
%{\cal A}_h = e^{i {\varphi}(h) }\;. \label{Weyl}
%\end{equation}
Using the relation $e^A \; e^B = \; e^{ A+B +\frac{1}{2}[A,B]}$,
%\begin{equation}
%e^A \; e^B = \; e^{ A+B +\frac{1}{2}[A,B]}, \label{exp_AB}
%\end{equation} 
valid for two operators $(A,B)$ commuting with $[A,B]$, one finds that the Weyl operators give rise to the following algebraic structure:
\begin{eqnarray}
	{\cal A}_h \;{\cal A}_h' & =  & e^{- \frac{1}{2} [{\varphi}(h), {\varphi}(h')] }\;{\cal A}_{(h+h')} = e^{ - \frac{i}{2} \Delta_{\textrm{PJ}}(h,h')}\;{\cal A}_{(h+h')},  \nonumber \\
	{\cal A}^{\dagger}_h & = & {\cal A}_{(-h)}, \label{algebra} 
\end{eqnarray} 
where $\Delta_{\textrm{PJ}}(h,h')$ is the smeared causal Pauli-Jordan expression \eqref{mint}. Setting $\varphi(h) = a_h + a^\dagger_h \;,$
%\begin{equation} 
%\varphi(h) = a_h + a^\dagger_h \;, \label{dec}
%\end{equation}
with $(a_h,a^\dagger_h)$ being the smeared annihilation and creation operators 
\begin{equation}
	a_h = \int \! \frac{d k}{2 \pi} \frac{1}{2 \omega_k} h^{*}(\omega_k,k) a_k \;, \qquad a_h^\dagger = \int \! \frac{d k}{2 \pi} \frac{1}{2 \omega_k} h(\omega_k,k) a^\dagger _k \;, \qquad [a_h,a^{\dagger}_{h'}]= \langle h | h'\rangle \;, \label{smaad}
\end{equation}
and $h(\omega_k,k)$ being the Fourier transformation of the test function $h(x)$, it follows that the vacuum expectation value of the operator ${\cal A}_h$ turns out to be
\begin{equation} 
	\langle 0 \vert  {\cal A}_h  \vert 0 \rangle = \; e^{-\frac{1}{2} {\lVert h\rVert}^2}, \label{vA}
\end{equation} 
where ${\lVert h\rVert}^2 \equiv \langle h | h \rangle$ and the vacuum state $\vert 0 \rangle$ is defined by $a_k \vert 0 \rangle=0, \forall k$. In particular, if $supp_f$ and $supp_g$ are spacelike separated, causality ensures that the Pauli-Jordan function vanishes. Thus, from the above properties, it follows the useful relation 
\begin{equation} 
	\langle 0 \vert  {\cal A}_f {\cal A}_{g}  \vert 0 \rangle =  \langle 0 \vert {\cal A}_{(f + g)} \vert 0 \rangle =
	\; e^{-\frac{1}{2} {\lVert f+g \rVert}^2}. \label{vAhh}
\end{equation}
A very important property of the von Neumann algebra ${\cal W}({\cal M})$, Eq.~\eqref{vn}, generated by the Weyl operators is that, due to the Reeh-Schlieder theorem \cite{Haag:1992hx,Witten:2018zxz}, the vacuum state $|0\rangle$ is both cyclic and separating, meaning that: $i)$ the set of states $\{{\cal A}_h \vert 0 \rangle$, $ {\cal A}_h \in {\cal W}({\cal M})\}$ are {\it dense} in the Hilbert space; $ii)$ the condition $\{{\cal A}_h \vert 0 \rangle = 0, {\cal A}_h \in {\cal W}({\cal M})\}$, implies ${\cal A}_h = 0$. \\\\In such a situation, one can apply the modular theory of Tomita-Takesaki \cite{Summers:1987fn,Summers:1987squ,Summers:1987ze,Guimaraes:2024mmp,Witten:2018zxz}, which will be presented for a generic von Neumann algebra ${\cal W}({\cal M})$ with a cyclic and separating vector state $|\omega\rangle$. To begin, it is helpful to remind the  notion  of {\it commutant} 
${\cal W'}({\cal M})$ of the  von Neumann algebra ${\cal W}({\cal M})$, namely   
\begin{equation} 
	{\cal W'}({\cal M}) = \{ w';  \;\;w'\;w = w \;w'\; \;\; \forall \; w \in {\cal W}({\cal M}) \}, \label{comm} 
\end{equation} 
{\it i.e.}, ${\cal W'}({\cal M})$ contains all elements which commute with each element of ${\cal W}({\cal M})$. Let us also state the  so-called Haag's duality \cite{Witten:2018zxz,Haag:1992hx,eckmann1973application}, ${\cal W'}({\cal M}) = {\cal W}({\cal M'}),$
%\begin{equation}
%	{\cal W'}({\cal M}) = {\cal W}({\cal M'}), \label{Hd}
%\end{equation}
namely, the commutant ${\cal W'}({\cal M})$ coincides with the elements of ${\cal W}({\cal M'})$ obtained by taking elements belonging to the symplectic complement ${\cal M'}$ of ${\cal M}$. This duality, to our knowledge, has only been proven in the case of free Bose fields \cite{eckmann1973application}. \\\\The 
Tomita-Takesaki construction makes use of  an anti-linear unbounded operator $S$ \cite{Bratteli:1979tw} acting on the von Neumann algebra ${\cal W}({\cal M})$ as 
\begin{align} 
	S \; w \vert \omega \rangle = w^{\dagger} \vert \omega \rangle, \qquad \forall w \in {\cal W}({\cal M})\;,  \label{TT1}
\end{align}  
from which it follows that $S \vert \omega \rangle = \vert \omega \rangle$ and $S^2 = 1$. Making use of the polar decomposition \cite{Bratteli:1979tw}
\begin{equation} 
	S = J \; \Delta^{1/2} \;, \label{PD}
\end{equation} 
where $J$ is the anti-linear modular conjugation operator and $\Delta$ is the self-adjoint and positive modular operator, the following properties holds \cite{Bratteli:1979tw} 
\begin{eqnarray} 
	J^{\dagger} & = & J \;, \qquad J^2 = 1 \;, \nonumber \\
	J \Delta^{1/2} \; J & = & \Delta^{-1/2} \;, \nonumber \\
	S^{\dagger} & = & J \Delta^{-1/2} \;, \qquad S^{\dagger} S^{\dagger} = 1 \;, \nonumber \\
	\Delta & = & S^{\dagger} S \;, \qquad \Delta^{-1} = S S^{\dagger} \;. \label{TTP}
\end{eqnarray}
We can now state the renowned  Tomita-Takesaki theorem \cite{Bratteli:1979tw}, namely:\\\ $i)$ $J \;{\cal W}({\cal M})\; J  =  {\cal W'}({\cal M})$ as well as $J \;{\cal W'}({\cal M}) \;J = {\cal W}({\cal M})$;\\\ $ii)$ there is  a one-parameter family of operators $\Delta^{it}$ with $t \in \mathbb{R}$ which leave ${\cal W}({\cal M})$ invariant, that is:
\begin{equation*}
	\Delta^{it} \; \;{\cal W}({\cal M}) \; \Delta^{-it} = \;{\cal W}({\cal M}).
\end{equation*}
This theorem has far reaching consequences and finds applications in many areas \cite{Summers:2003tf}. As far as the Bell-CHSH inequality is concerned, it provides a powerful way of obtaining Bob's operators from Alice's ones by means of the anti-unitary operator $J$. The construction goes as follows: One picks up two test functions $(f,f')\in {\cal M}({\cal O})$ and consider  the two Alice's operators $({\cal A}_f, {\cal A}_{f'})$,
\begin{equation} 
	{\cal A}_f = e^{i {\varphi}(f) } \;, \qquad  {\cal A}_{f'} = e^{i {\varphi}(f') } \;. \label{AAop}
\end{equation} 
Thus, for Bob's operators, we write 
\begin{equation} 
	J {\cal A}_f J \;, \qquad J {\cal A}_{f'} J \;. \label{Bobbop}
\end{equation} 
From the Tomita-Takesaki Theorem, it follows that the two set of operators $({\cal A}_f, A_{f'})$ and $(J{\cal A}_f J, J {\cal A}_{f'}J)$ fulfill the necessary requirements, since $(J{\cal A}_f J, J {\cal A}_{f'}J)$ belong to the commutant ${\cal W'}({\cal M})$. Also, it is worth underlining that  the action of the operators $J$ and $\Delta$ may be lifted directly into the space of the test functions  \cite{eckmann1973application,rieffel1977bounded,Guido:2008jk}, giving rise to an analogue of the Tomita-Takesaki construction, namely,  
\begin{equation} 
	J {\cal A}_f J =  J e^{i {\varphi}(f) } J  \equiv e^{-i {\varphi}(jf) } \;, \label{jop}
\end{equation} 
where the test function $jf \in {\cal M'}$. Analogously, 
\begin{equation} 
	\Delta^{1/2} {\cal A}_f  \Delta^{-1/2} \equiv e^{i {\varphi}(\delta^{1/2}f) } \;. \label{dp}
\end{equation} 
The operators $(j,\delta)$ are such that \cite{eckmann1973application,rieffel1977bounded,Summers:1987squ,Guido:2008jk}
\begin{eqnarray}
	s & = & j \delta^{1/2} \;, \qquad s^2=1 \;, \nonumber \\
	j \delta^{1/2} j & = & \delta^{-1/2} \;, \nonumber  \\
	s^{\dagger} & = & j \delta^{-1/2} \;, \qquad s^{\dagger} s^{\dagger} = 1 \;. \label{ssd}
\end{eqnarray}
The operator $j$ is anti-unitary, while $\delta$ is self-adjoint and positive. Moreover, from \cite{Summers:1987squ}, one learns that, in the case in which the region ${\cal O}$ is a wedge region of the Minkowski spacetime, the spectrum of the operator $\delta$ is the whole positive real line, {\it i.e.} $\log(\delta) =\mathbb{R}$. This result follows from the analysis of Bisognano-Wichmann \cite{Bisognano:1975ih} of the Tomita-Takesaki modular operator for wedge regions $({\cal O}_R,{\cal O}_L)$ 
\begin{equation}
	{\cal O}_R =\left\{ (x,t)\;, x \ge |t| \right\} \;, \qquad  {\cal O}_L =\left\{ (x,t)\;, -x \ge |t| \right\} \;. \label{wedges}
\end{equation}
The wedge regions $({\cal O}_R,{\cal O}_L)$ are causal complement of each other and are left invariant by Lorentz boosts. In what follows, we shall always consider the region ${\cal O} $ to be a wedge region. For instance, one may figure out that Alice is located in the right  wedge ${\cal O}_R$, while Bob is in the left one ${\cal O}_L$. \\\\The operators $(s,s^{\dagger})$ have the physical meaning of projecting into the space ${\cal M}$ and its symplectic complement ${\cal M'}$, namely, one can show \cite{rieffel1977bounded, Summers:1987squ,Guido:2008jk}  that  a test function $f$ belongs to $\cal M$ if and only if
\begin{equation} 
	s f = f \;. \label{sf}
\end{equation} 
Analogously, a test function $g$ belongs to the symplectic complement ${\cal M'}$ if and only if
\begin{equation} 
	s^{\dagger} g = g \;. \label{sdg}
\end{equation} 
These properties enable us to construct a useful set of test functions for Alice and Bob. Following \cite{Summers:1987fn,Summers:1987squ}, Alice's test functions $(f,f')$ can be  specified as follows: Picking up the spectral subspace of $\delta$ specified by $[\lambda^2-\varepsilon, \lambda^2+\varepsilon ] \subset (0,1)$ and introducing  the normalized vector $\phi$ belonging to this subspace, one writes  
\begin{equation}
	f  = \eta  (1+s) \phi \;, \qquad f' = \eta' (1+s) i \phi \;, 
	\label{nmf}
\end{equation}
where $(\eta, \eta')$ are free arbitrary parameters, corresponding to the norms of $(f,f')$.  According to the setup outlined above, Eq.~\eqref{nmf} ensures that $s f = f $ and $s f'= f' \;.$
%\begin{equation}
%s f = f  \;, \qquad s f'= f' \;.  \label{fafa}
%\end{equation}
Moreover, one checks  that $j\phi$ is orthogonal to $\phi$, {\it i.e.} $\langle \phi |  j\phi \rangle = 0$. In fact, from 
\begin{align} 
	\delta^{-1} (j \phi) =  j (j \delta^{-1} j) \phi = j (\delta \phi), 
	\label{orth}
\end{align}
it follows that the modular conjugation $j$ exchanges the spectral subspace $[\lambda^2-\varepsilon, \lambda^2+\varepsilon ]$ into $[1/\lambda^2-\varepsilon,1/ \lambda^2+\varepsilon ]$, ensuring that $\phi$ and $j \phi$ are orthogonal.   

Concerning now the pair of Bob's test functions, they are given by  $( \sigma j(1+s)\phi, - \sigma'j(1+s)i\phi)$, with $(\sigma,\sigma')$ normalization factors. In fact, we have 
\begin{equation} 
	s^{\dagger}(jf) = jf \;, \qquad s^{\dagger}(jf') = jf'  \;, \label{jffp}
\end{equation}
meaning that, as required by the relativistic causality, $(jf,jf')$ belong to the symplectic  complement $\mathcal{M'}(\mathcal{O})$.  \\\\Finally, taking into account that $\phi$ belongs to the spectral subspace $[\lambda^2-\varepsilon, \lambda^2+\varepsilon ] $, it follows that \cite{Summers:1987fn,Summers:1987squ,Guimaraes:2024mmp}, in the limit $\varepsilon \rightarrow 0$,
\begin{align}
	\vert\vert f \vert\vert^2  &= \vert\vert jf  \vert\vert^2 = \eta^2 (1+\lambda^2) \;, \nonumber \\
	\langle f \vert jf  \rangle &= 2 \eta^2 \lambda \;,  \nonumber \\
	\vert\vert f' \vert\vert^2  &= \vert\vert jf'  \vert\vert^2 = {\eta'}^2 (1+\lambda^2) \;, \nonumber \\
	\langle f' \vert  jf' \rangle &= 2 {\eta'}^2 \lambda \;,  \nonumber \\
	\langle f \vert f'\rangle & = i \eta \eta'(1-\lambda^2) \;, \nonumber \\
	\langle f \vert jf' \rangle &=  0  \;.  \label{ssfl}
\end{align}
\vspace{-1.4cm}

\subsection{The violation of the Bell-CHSH inequality in Quantum Field Theory} 

\begin{figure}[t!]
	\begin{minipage}[b]{0.4\linewidth}
		\includegraphics[width=\textwidth]{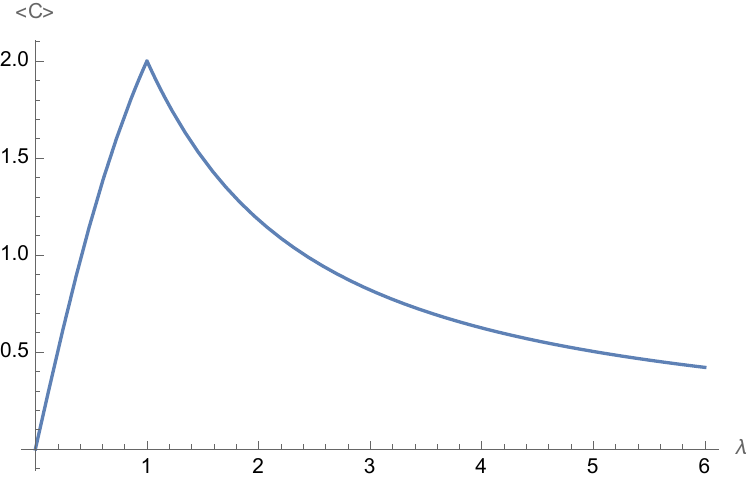}
	\end{minipage} \hfill
	\caption{Behavior of the Bell-CHSH correlator  $\langle {\cal C} \rangle$ as a function of the parameters $\lambda$. Although $\lambda$ belongs to the spectral interval $[0,1]$, the plot has been extended over a larger interval to show that $\langle {\cal C} \rangle$ is bounded by 2. }
	\label{plot}
\end{figure}

We have now all tools to discuss the violation of the Bell-CHSH inequality in QFT. As already mentioned, we consider the following expressions 
\begin{eqnarray} 
	{\cal P}_f &  = & {\it sign}(\varphi(f)) = \frac{2}{\pi} \int_{0}^{\infty} \frac{dk}{k}\; \sin(k \varphi(f) ) \;, \qquad 
	{\cal P}_{f'}   =  {\it sign}(\varphi(f')) = \frac{2}{\pi} \int_{0}^{\infty} \frac{dk}{k}\; \sin(k \varphi(f') ) \;, \nonumber \\
	{\cal P}_g &  = & {\it sign}(\varphi(g)) = \frac{2}{\pi} \int_{0}^{\infty} \frac{dk}{k}\; \sin(k \varphi(g) ) \;, \qquad 
	{\cal P}_{g'}   =  {\it sign}(\varphi(g')) = \frac{2}{\pi} \int_{0}^{\infty} \frac{dk}{k}\; \sin(k \varphi(g') ) \;. \label{ss1}
\end{eqnarray}
These operators fulfill the required algebraic conditions \eqref{c1} and \eqref{c2}. Let us proceed with the evaluation of the correlation function $\langle 0| {\cal P}_f \;{\cal P}_g |0\rangle$. Making use of the properties of the Weyl operators, one finds 
\begin{eqnarray} 
	\langle 0| {\cal P}_f \;{\cal P}_g |0\rangle & = & - \frac{1}{\pi^2} \int_0^\infty \frac{dk dq}{kq}\; \langle 0| ( e^{i k \varphi(f)} - e^{-i k \varphi(f)}) ( e^{i q \varphi(g)} - e^{-i k \varphi(g)}) |0\rangle \nonumber \\
	& = & \frac{4}{\pi^2} \int_0^\infty \frac{dk dq}{kq}\; e^{-\frac{1}{2}(k^2 ||f||^2 + q^2 ||g||^2)} \sinh( kq \langle f| g \rangle) \nonumber \\
	& = & \frac{4}{\pi^2} \int_0^\infty \frac{dk dq}{kq}\; e^{-\frac{k^2}{4}} e^{-\frac{q^2}{4}} \sinh( \frac{kq \;\langle f| g \rangle}{2\; ||f||\; ||g||}) \;. \label{ss2}
\end{eqnarray}
Recalling that the imaginary error function ${\it erfi}(w)$ can be represented as 
\begin{equation} 
	{\it erfi}(w) = \frac{2}{\pi} \int_0^\infty \frac{dt}{t} e^{-\frac{t^2}{4}} \; \sinh(w t) \;, \label{erfi} 
\end{equation}  
it follows that 
\begin{equation} 
	\langle 0| {\cal P}_f \;{\cal P}_g |0\rangle = \frac{2}{\pi} \int_0^\infty \frac{dk}{k} e^{-\frac{k^2}{4}}\; {\it erfi}\left( \frac{k \;\langle f| g \rangle}{2\; ||f||\; ||g||}\right) = \frac{2}{\pi} \arcsin\left( \frac{\langle f| g \rangle}{||f||\; ||g||}\right) \;. \label{ss3}
\end{equation}
Therefore, for the Bell-CHSH correlator in the vacuum state, we get 
\begin{eqnarray} 
	\langle 0 |\; {\cal C}\;| 0 \rangle &=&  \langle 0| {\cal P}_f \;{\cal P}_g  + {\cal P}_{f'} \;{\cal P}_g + {\cal P}_f \;{\cal P}_{g'} - {\cal P}_{f'} \;{\cal P}_{g'} |0\rangle \nonumber \\
	&=&\frac{2}{\pi}\left(  \arcsin\left( \frac{\langle f| g \rangle}{||f||\; ||g||}\right)+  \arcsin\left( \frac{\langle f'| g \rangle}{||f'||\; ||g||}\right) +  \arcsin\left( \frac{\langle f| g' \rangle}{||f||\; ||g'||}\right) -  \arcsin\left( \frac{\langle f'| g' \rangle}{||f'||\; ||g'||}\right)\right) \;. \label{ss4}
\end{eqnarray} 
Finally, employing expressions  \eqref{ssfl}, it turns out that the Bell-CHSH correlator can be cast in closed form as 
\begin{equation} 
	\langle 0 |\; {\cal C}\;| 0 \rangle = \frac{4}{\pi}  \arcsin\left( \frac{2 \lambda}{1+\lambda^2}\right) \;. \label{cls}
\end{equation} 
Unfortunately, this expression is bounded by the classical value 2, as it can be seen from Fig.~\eqref{plot}. 
However, the situation may be  changed by employing unitary transformations, namely 
\begin{eqnarray} 
	{\cal P}_{f} & \rightarrow & {\hat {\cal P}}_f = {\cal U}^{\dagger}_A \; {\cal P}_f \;{\cal U}_A = e^{i {\hat\alpha} \varphi(f')  } \; {\cal P}_f \; e^{-i{\hat  \alpha} \varphi(f') } \;, \nonumber \\
	{\cal P}_{f'} & \rightarrow & {\hat {\cal P}}_{f'} = {\cal U'}^{\dagger}_A \; {\cal P}_{f'} \;{\cal U'}_A = e^{i{\hat \alpha}' \varphi(f)  } \; {\cal P}_{f'} \; e^{-i {\hat \alpha}' \varphi(f) } \;, \nonumber \\
	{\cal P}_{g} & \rightarrow & {\hat {\cal P}}_{g} = {\cal U}^{\dagger}_B \; {\cal P}_{g} \;{\cal U}_B = e^{i{\hat \beta} \varphi(g')  } \; {\cal P}_{g} \; e^{-i {\hat \beta} \varphi(g') } \;, \nonumber \\
	{\cal P}_{g'} & \rightarrow & {\hat {\cal P}}_{g'} = {\cal U'}^{\dagger}_B \; {\cal P}_{g'} \;{\cal U'}_B = e^{i{\hat \beta}' \varphi(g)  } \; {\cal P}_{g'} \; e^{-i {\hat \beta}' \varphi(g) } \;,  \label{unitqft}
\end{eqnarray}
where $({\hat \alpha}, {\hat \beta}, {\hat \alpha'}, {\hat \beta'})$ are free arbitrary parameters. By construction, the operators $({\hat {\cal P}}_f,{\hat {\cal P}}_{f'},{\hat {\cal P}}_{g} ,{\hat {\cal P}}_{g'} )$ fulfill the conditions \eqref{c1} and \eqref{c2}. Moreover, using the relations between the Weyl operators, it follows that the effect of performing a unitary transformation is equivalent to a shift in the field, namely 
\begin{equation} 
	{\hat {\cal P}}_{f} = {\it sign}(\varphi(f) + \alpha) =  \frac{2}{\pi} \int_{0}^{\infty} \frac{dk}{k}\; \sin(k \varphi(f) + k \alpha) \;, \label{ss5}
\end{equation}
where, from Eqs.~\eqref{ssfl}, the parameter $\alpha$ is related to ${\hat \alpha}$ by 
\begin{equation}
	\alpha = {\hat \alpha} \;(2 \eta \eta' (1 -\lambda^2)) \;. \label {redef}
\end{equation}
Similar equations hold for the other operators $({\hat {\cal P}}_{f'},{\hat {\cal P}}_{g} ,{\hat {\cal P}}_{g'} )$. \\\\Evaluating now the   Bell-CHSH correlator for the transformed operators, one obtains 
\begin{align} 
	\langle 0 |\; {\hat {\cal C}}\;| 0 \rangle &=  \langle 0| {\hat {\cal P}}_f \;{\hat {\cal P}}_g  + {\hat {\cal P}}_{f'} \;{\hat {\cal P}}_g + {\hat {\cal P}}_f \;{\hat {\cal P}}_{g'} -{\hat  {\cal P}}_{f'} \;{\hat {\cal P}}_{g'} |0\rangle \nonumber \\
	&= - \frac{2}{\pi^2} \int_0^\infty \frac{dk dq}{k q} \left( e^{-\frac{1}{2}( k^2 \eta^2 (1+\lambda^2) + q^2\sigma^2(1+\lambda^2) + 4kq \eta \sigma \lambda) }  \cos(\alpha k + \beta q) - e^{-\frac{1}{2}( k^2 \eta^2 (1+\lambda^2) + q^2\sigma^2(1+\lambda^2) - 4kq \eta \sigma \lambda) }  \cos(\alpha k - \beta q) \right) \nonumber \\
	& - \frac{2}{\pi^2} \int_0^\infty \frac{dk dq}{k q} \left( e^{-\frac{1}{2}( k^2 \eta'^2 (1+\lambda^2) + q^2\sigma^2(1+\lambda^2) ) }  \cos(\alpha' k + \beta q) - e^{-\frac{1}{2}( k^2 \eta'^2 (1+\lambda^2) + q^2\sigma^2(1+\lambda^2) ) }  \cos(\alpha' k - \beta q) \right) \nonumber \\
	& - \frac{2}{\pi^2} \int_0^\infty \frac{dk dq}{k q} \left( e^{-\frac{1}{2}( k^2 \eta^2 (1+\lambda^2) + q^2\sigma'^2(1+\lambda^2) ) }  \cos(\alpha k + \beta' q) - e^{-\frac{1}{2}( k^2 \eta^2 (1+\lambda^2) + q^2\sigma'^2(1+\lambda^2) ) }  \cos(\alpha k - \beta' q) \right) \nonumber \\
	&-  \frac{2}{\pi^2} \int_0^\infty \frac{dk dq}{k q} \Big( e^{-\frac{1}{2}( k^2 \eta'^2 (1+\lambda^2) + q^2\sigma'^2(1+\lambda^2) - 4kq \eta' \sigma' \lambda) }  \cos(\alpha' k + \beta' q)\nonumber\\ 
	&- e^{-\frac{1}{2}( k^2 \eta'^2 (1+\lambda^2) + q^2\sigma'^2(1+\lambda^2) + 4kq \eta' \sigma' \lambda) }  \cos(\alpha' k - \beta' q) \Big) 
	\label{nnb}
\end{align}
Unlike Eq.~\eqref{cls}, the above expression contains the four arbitrary parameters $(\alpha, \alpha', \beta, \beta')$ which can be chosen so as to violate the Bell-CHSH inequality. For instance, setting 
\begin{eqnarray}
	\eta & = & 0.00132432, \qquad \eta' = 1.04948, \qquad  \lambda = 0.811798, \qquad \alpha  = -12.6591, \nonumber \\
	\alpha' & =  & -8.04613, \qquad  \beta  =-6.00724, \qquad  \beta' =-6.05653, \nonumber \\
	\sigma & = & 0.0737113, \qquad  \sigma'= 4.8949, \; 
	\label{values}
\end{eqnarray}
it turns out that 
\begin{equation} 
	| \langle 0 |\; {\hat {\cal C}}\;| 0 \rangle| =  2.02034 \;, \label{vv}
\end{equation} 
exhibiting a violation of the Bell-CHSH inequality.
%The double integrals in expression \eqref{nnb} have been evaluated using the \texttt{QuasiMonteCarlo} method from Mathematica, with precision set by Maxpoints = $10^9$. Concerning the parameters $(\eta, \eta', \sigma, \sigma', \lambda, \alpha, \alpha', \beta, \beta')$ we have performed random tests. Each test has been repeated $10^5$ times. At each run, we have obtained a rather large number of values yielding to violations similar to that reported in eq.\eqref{vv}, showing thus the usefulness of the unitary transformations in the study of the Quantum Field Theory version of the Bell-CHSH inequality. 
The double integrals appearing in expression \eqref{nnb} were evaluated numerically using Mathematica’s \texttt{QuasiMonteCarlo} routine, with the sampling precision controlled by setting \texttt{MaxPoints} up to $10^9$. The parameter space $(\eta, \eta', \sigma, \sigma', \lambda, \alpha, \alpha', \beta, \beta')$ was explored through randomized trials, each of which was repeated $10^5$ times. For essentially every trial, a substantial subset of sampled configurations produced violations of the type displayed in Eq. \eqref{vv}, thereby underscoring the operational relevance of the unitary transformations in analyzing the Quantum Field Theory formulation of the Bell-CHSH inequality. The computational cost of the integration grows linearly with the number of sampling points, which precludes increasing \texttt{MaxPoints} without bound. Empirically, $10^9$ sampling points constituted a practical upper limit for the present computations. Within this range, we identified the minimal value of \texttt{MaxPoints} required to guarantee convergence of the integral to within a prescribed \texttt{PrecisionGoal}.

%%%%%%%%%%%%%%%%%%%%%%%%%%%%%%%%%%%%%%%
\section{The Proca field in ($1+1$)-dimensions}\label{ProcaField}
We now turn to the case of the massive spin-1 Proca field $V_\mu$ in two-dimensional Minkowski spacetime. The dynamics of the field $V_{\mu}$ is described by the Lagrangian density
\begin{equation}
	\mathcal{L} = - \frac{1}{4} F_{\mu \nu}(V)F^{\mu \nu}(V) + \frac{M^2}{2} V^{\mu}V_{\mu}\;,
\end{equation}
where $F_{\mu \nu}(V) = \partial_{\mu} V_{\nu} - \partial_{\nu} V_{\mu}\;$ and the parameter 
$M$ corresponds to the physical mass of the spin-1 vector field. The quantized field admits the plane-wave mode expansion
\begin{equation}
	V_{\mu}(t,x) = \int \frac{dk}{2 \pi} \frac{1}{2 \omega_k}[\epsilon_{\mu}(k)a(k)e^{-ik_{\mu}x^{\mu}} + \epsilon_{\mu}(k)a^{\dagger}(k)e^{ik_{\mu}x^{\mu}}]\;,
\end{equation}
with dispersion relation $\omega_k=\sqrt{k^2+M^2}$.
The polarization vectors satisfy the standard conditions
\begin{equation}
	k^\mu \,\epsilon_\mu(k) = 0, \qquad
	\epsilon^\mu(k)\,\epsilon_\mu(k) = -1, \qquad
	\epsilon_\mu(k)\,\epsilon_\nu(k) = -\left(\eta_{\mu\nu}-\frac{k_\mu k_\nu}{M^2}\right)\;,
\end{equation}
with $\eta_{\mu\nu} = \textrm{diag}(+,-)$. The creation and annihilation operators fulfill the canonical commutation relations
\begin{equation}
	\begin{aligned}
		[a_k, a^{\dagger}_{q}] &= (2\pi)\,2\omega_k\,\delta(k-q),\\
		[a_k, a_{q}] &= [a^{\dagger}_k, a^{\dagger}_{q}] = 0\;.
	\end{aligned}
\end{equation}

A distinctive property of the Proca field is its transversality condition $\partial^{\mu}V_{\mu} = 0,$ which implies that only transverse test functions contribute when the field is smeared. For a smooth compactly supported function $f_\mu(\vb*{x})$, we define
\begin{equation}
	V(f)= \int d^2\vb*{x} \,V_{\mu}(\vb*{x}) f^{\mu}(\vb*{x})=a_f + a^{\dagger}_f,
\end{equation}
with the inner product
\begin{equation}
	[a_f,a^{\dagger}_g] = \langle f \vert g \rangle_V = -\int \frac{dk}{2 \pi} \frac{1}{2 \omega_k} f^{*\,\mu}(k)g^{\nu}(k) \left(\eta_{\mu\nu}-\frac{k_\mu k_\nu}{M^2}\right)\;.
\end{equation}
The Fourier transform of the test functions is given by
\begin{equation}
	f^{\mu}(k) = \int d^2\vb*{x} \,e^{-ik_{\mu}x^{\mu}} f^{\mu}(\vb*{x})\;.
\end{equation}
Equivalently, in configuration space, one finds
\begin{equation}
	\langle f \vert g \rangle_{V} = - \int d^2\vb*{x}\,d^2\vb*{y}\, f^{\mu}(\vb*{x}) \left(\eta_{\mu\nu}-\frac{\partial^{\vb*{x}}_\mu \partial^{\vb*{x}}_\nu}{M^2}\right) \Delta(\vb*{x}-\vb*{y}) g^{\nu}(\vb*{y})\;,
\end{equation}
where
\begin{align}
	\Delta(\vb*{x}-\y) = \int \frac{dk}{2 \pi} \frac{1}{2 \omega_k} e^{-i k_{\mu} (x^{\mu}-y^{\mu})} 
	= H(\x-\y) + \frac{i}{2} \Delta_{\rm PJ}(\x-\y)\;,
\end{align}
with $H(\x-\y)$ and $\Delta_{\rm PJ}(\x-\y)$ denoting the Hadamard and Pauli–Jordan distributions, respectively [cf. Eqs.~\eqref{InnerProduct}--\eqref{PJH}].

To analyze the consequences of transversality, let an arbitrary test function $h_{\mu}(\x)$ be decomposed into transverse and longitudinal parts,
\begin{equation}
	h_{\mu} = h_{\mu} ^T + h_{\mu} ^L, \qquad h_{\mu} ^T = \left(\eta_{\mu\nu}-\frac{\partial_\mu \partial_\nu} {\partial^2}\right) h^{\nu}, \qquad
	h_{\mu} ^L = \frac{\partial_\mu \partial_\nu}{\partial^2} h^{\nu}\;,
\end{equation}
where $\partial^2=\partial_0^2-\partial_1^2.$ Then, the longitudinal component does not contribute to the smeared field:
%\begin{align}
%   \int d^2x\, V^{\mu} h^{L}_{\mu} &= \int d^2x \, V^{\mu} \frac{\partial_\mu \partial_\nu}{\partial^2} h^{\nu} = 0\;,
%\end{align}
\begin{equation}
	V(h) = \int d^2\x\, V^{\mu} h_{\mu} =\int d^2\x\, V^{\mu} h^{T}_{\mu} = V(h^{T})\;,
\end{equation}
which shows that only transverse test functions are physically relevant. A convenient realization is
\begin{equation}
	f_{\mu}(t,x) = \frac{1}{M} (\partial_1 f, \partial_0 f)\;,
\end{equation}
with $f$ a smooth scalar test function. This ensures that $\partial^{\mu} f_{\mu} = \dfrac{1}{M}  (\partial_0 \partial_1 f - \partial_1 \partial_0 f ) = 0.$ Substituting into the inner product yields
\begin{align}
	\langle f \vert g \rangle_{V} &= - \int d^2\x \, d^2\y \, f^{\mu}(\x) g_{\mu}(\y) \Delta(\x-\y) \nonumber \\
	&= - \frac{1}{M^2}\int d^2\x \, d^2\y \, (\partial_1^{\x} f \partial_1^{\y} g - \partial_0^{\x} f \partial_0^{\y} g)\Delta(\x-\y) \nonumber \\
	&= - \frac{1}{M^2}\int d^2\x \, d^2\y \, f(\x) g(\y) (\partial_1^{\x}  \partial_1^{\y}  - \partial_0^{\x}  \partial_0^{\y})\Delta(\x-\y) \nonumber \\
	&= \frac{1}{M^2}\int d^2\x \, d^2\y \, f(\x) g(\y) (-\partial_0 ^2 + \partial_x ^2) \Delta(\x-\y)\;.
\end{align}
Using the Klein–Gordon equation $(\partial_0 ^2 - \partial_x ^2 + M^2) \Delta(\x-\y)=0$, this reduces precisely to the scalar-field inner product,
\begin{align}
	\langle f \vert g \rangle_V &= \int d^2\x \, d^2\y \, f(\x) \Delta(\x-\y) g(\y) = \langle f \vert g \rangle\;,
	\label{Innerfg_V_fg}
\end{align}
demonstrating that, in two spacetime dimensions, the massive Proca field is dynamically equivalent to a scalar field. The reduction follows from the fact that a vector field in $(1+1)$-dimensions carries a single physical degree of freedom. Consequently, the analysis of Bell-CHSH violations parallels the scalar case,. Thus, in $(1+1)$-dimensions, the Proca field exhibits no qualitative distinction from the massive scalar field regarding Bell-CHSH violations as expected from their duality relation: both theories possess the same physical content, leading to identical Bell-CHSH violations in the vacuum and exhibiting the same enhancement under unitary deformations.

%Thus, in $(1+1)$-dimensions, the Proca field exhibits no qualitative distinction from the massive scalar field regarding Bell-CHSH violations: both theories display identical entanglement structures in the vacuum, and both benefit equally from the introduction of unitary deformations.

\section{Conclusions}\label{Cc}
We have presented an explicit QFT realization of Bell-CHSH tests in which unitary deformations of modular-localized observables serve as tunable parameters for enhancing nonlocal correlations in the vacuum. Using the bounded Hermitian operator ${\it sign}(\varphi(f))$ and the Tomita-Takesaki construction to ensure strict spacelike separation, we confirmed that the undeformed correlator saturates only the classical bound, while unitary shifts induce genuine quantum violations. Although the resulting enhancement is quantitatively mild, several structural features of the modular setup help explain this behavior: the strong constraints imposed by wedge localization, the limited functional flexibility of operators built from a single smeared field, and the highly nontrivial geometry of the modular spectral subspaces. These considerations suggest that larger violations may require either more intricate bounded functionals of the field or multi-operator architectures shaped by modular flow. Nevertheless, the present approach remains compelling, as it provides a fully relativistic, computationally explicit demonstration of how unitary deformations alter QFT Bell tests, and it offers a principled framework for systematically improving Bell operators in future studies. The mechanism persists for the Proca field in $1+1$ dimensions via its dual scalar description, reinforcing the broader relevance of our construction for probing vacuum entanglement in relativistic quantum systems.

Several directions for future work naturally follow. Extending the present framework to higher-dimensional spacetimes would allow modular operators with richer spectra, potentially offering new handles for optimizing Bell observables. It would also be valuable to explore interacting field theories, where vacuum correlations reflect nontrivial dynamical structure, and to analyze gauge-invariant sectors in both Abelian and non-Abelian models, where additional constraints may qualitatively alter the attainable violations.

\section*{Acknowledgments}
The authors would like to thank the Brazilian agencies CNPq, CAPES end FAPERJ for financial support.  S.~P.~Sorella, I.~Roditi, and M.~S.~Guimaraes are CNPq researchers under contracts 301030/2019-7, 311876/2021-8, and 309793/2023-8, respectively. A.~F.~Vieira is supported by a postdoctoral grant from FAPERJ in the Pós-doutorado Nota 10 program,
grant No. E-26/200.135/2025. F.~M.~Guedes acknowledges FAPERJ for financial support under the contract SEI-260003/008326/2025. D.~O.~R.~Azevedo acknowledges PEDECIBA for financial support.

%\end{acknowledgments}

%\appendix

%%%%%%%%%%%%%%%%%%%%%%%%%%%%%%%%%%%%%%%%%%%%%%%%%%%%%%%%%%%%%%%%%%%%%%%%%%%%%%%%%%%%%%%%%%%%%%%%%%%%%%%%%%
	
	\bibliographystyle{unsrt}
	\bibliography{Refs}
\end{document}